\input amstex
\documentstyle{amsppt}  \magnification=\magstephalf \TagsOnRight \NoRunningHeads \nologo
\vsize 23.5 truecm \hsize=16 truecm 
\parskip=2pt

\def\={\;=\;} \def\+{\,+\,}  \def\h{\frac12}
\def\a{\alpha} \def\b{\beta}  \def\l{\lambda} \def\g{\gamma} \def\k{\kappa} 
\def\G{\Gamma}   \def\L{\Lambda}  \def\tf{\tilde f}
\def\Z{\Bbb Z}  \def\R{\Bbb R}   \def\C{\Bbb  C} \def\H{\Bbb H}
\def\hq{{\hat q}}  \def\abcd{\bigl(\smallmatrix  a&b\\c&d\endsmallmatrix\bigr)}
    
 \def\nl#1{\newline\noindent{\rm(#1)}}


  \def\dual{1}  \def\s{2}  \def\pih{3}  \def\pihint{4}  \def\de0{5}   \def\hvb{7}
\def\deI{6} \def\pihvb{8} \def\pihvbint{9}  \def\deII{10}  \def\deIII{11} \def\ltau{12}  \def\th{13} 
\def\der{14} \def\ab{15}   \def\fIIint{16} \def\ffI{17}   \def\ffw{18}  \def\pif{19} 
\def\php{20}    \def\pihk{21}   \def\pihkint{22}  \def\pihpr{23}  \def\pihra{24}
  \def\modx{25} \def\nozs{26} \def\nogzs{27} \def \cblock{28}  \def\newi{29} \def\const{30} 
\def\thiii{31}  \def\vder{32}  \def\vint{33} \def\vintii{34}  \def\deIIIp{35} \def\pihvbnew{36}  \def\rect{37}
\def\hof{38}  \def\hoff{39}

\topmatter
\title Crossing Probabilities and Modular Forms\endtitle
\author Peter Kleban\footnote{LASST and Department of Physics \& Astronomy, University of Maine, Orono, Maine, USA
\hskip2.6cm\null} and Don Zagier  \footnote{Max-Planck-Institut f\"ur Mathematik, Bonn, Germany\hskip7.4cm\null}
\endauthor

\abstract   
We examine crossing probabilities and free energies for conformally invariant critical 2-D systems in
rectangular geometries, derived via conformal field theory and Stochastic L\"owner Evolution methods. 
These quantities are shown to exhibit interesting modular behavior, although the  physical meaning of 
modular transformations in this context is not clear.  We show that in many cases these functions are  
completely characterized by very simple transformation properties.  In particular, Cardy's function for the percolation crossing probability
(including the conformal dimension $1/3$), follows from a  simple modular argument. A new type of 
``higher-order modular form"  arises and its properties are discussed briefly. \endabstract 
\endtopmatter

\subheading{1. Introduction}
There are extensive and well-known applications of modular invariance to various quantities arising
in conformal field theory (CFT), beginning with the work of Cardy \cite{Ca1} on the implications of 
modular invariance for the operator content of a given theory.  However, in these cases the system is 
defined on a torus so that the modular invariance is implicit already from the definition of the problem.  
In this paper we examine the modular behavior of several quantities defined on compact two-dimensional 
figures with boundary, typically rectangles, where there is no obvious reason to expect modular properties.  
More explicitly, in the cases we consider, the modular operation $S$  corresponding to $\tau\to-1/\tau$ 
is implied by a symmetry of the problem, but the operation $T$  which takes $\tau$ to $\tau+1$ has
no apparent physical interpretation; its origin is mysterious.  Despite this, we are able display some 
new and interesting modular properties of known solutions to several problems from CFT   and the recently 
developed Stochastic L\"owner Evolution (SLE) method \cite{LSW1}.  Examination of these properties for 
the crossing probabilities (originally defined for percolation, but more recently extended to other 
processes via SLE) leads us to define a new type of modular quantity that we call an {\it $n^{th}$ 
order modular form}. Conversely, we show that by postulating a specific form for the crossing probability 
(and similarly for the partition function),  it is possible to reproduce the explicit forms for these quantities.

In Sections~2 and 3 we briefly review percolation and the exact analytic forms of the crossing probabilities 
given by CFT \cite{Ca2}, \cite{W} and transform these results into a form suitable for the present analysis. 
We also mention their generalization to other processes via SLE \cite{LSW1}.  Section~4 briefly recapitulates 
some of the basic properties of modular forms. In Section~5 we prove several theorems showing that the crossing
probabilities are uniquely characterized by two very simple assumptions, a symmetry property and an assumption 
about the form of their $q$-expansion.  In addition, we exhibit a possible generalization of one of the 
crossing probabilities to the SLE processes.    In Section~6, we consider the (universal part of the) partition 
function for systems on a rectangle and show that the exact expression, already known by a CFT calculation \cite{KlV}, 
has a modular characterization of a similar kind. Section~7 considers the ``higher order modular forms"   
that arise from the crossing probabilities.  Roughly speaking, a first order modular form is an ordinary 
modular form and an $n$th order modular form is a function whose deviation from modularity is an $(n-1)$st 
order modular form.  Section~8 concludes the paper with a brief discussion.

The subject we consider lies between mathematics and physics. Therefore some introductory matter from both 
fields has been included to make our treatment more comprehensible to those with background in one area 
but not the other.

\subheading{2. Crossing Probabilities}

In this section, we first briefly review percolation (for a more complete treatment, see \cite{Ke} or \cite{StA}),
and the definition of the crossing probabilities \cite{LPPS}.  Then we give the exact analytic forms for the latter
quantities obtained via boundary CFT (\cite{Ca2}, \cite{W}) and their generalization from SLE \cite{LSW1}.  The 
various crossing probabilities are displayed in three different ways:  as an (ordinary or generalized) hypergeometric
function of a certain cross-ratio $\l$ (to be defined);  as  the integral of an algebraic function of $\l$; and 
directly in terms of  the aspect ratio $r$ of the rectangle (following \cite{Zi1} and \cite{Zi2}).  In the last 
case, the probabilities may also be written as series in rational powers of the parameter $q = e^{2 \pi i \tau}$, 
where $\tau :=  i r$.

Percolation is perhaps the simplest non-trivial model in statistical mechanics. It is very easy to define, and 
exhibits a second-order phase transition between the percolating and non-percolating states. There are various 
types of percolation; for definiteness, we consider bond percolation on a   lattice with square unit mesh. The
system considered is a finite rectangular $L \times L'$ lattice where $L, L' \to \infty$ with fixed aspect 
ratio $r$ = width/height = $L/L'$. A bond is placed with (independent) probability $p$ on each edge of the 
lattice. Consequently, there are $2^N$ possible bond configurations with $0 \le N_B\le N$, where $N_B$ is the 
number of bonds in a given configuration and $N$ is the total number of edges. The connected bonds in each 
configuration form clusters.  Note that for any configuration, either there is a cluster spanning the lattice 
from left to right, in which case the dual lattice has no vertical spanning cluster, or there is no   horizontal
spanning cluster on the lattice and the dual lattice has a vertical spanning cluster. For $p > p_c$, as the 
lattice is taken to infinity, an infinite cluster appears.  For $p \le p_c$, there is no infinite cluster.  
In the case at hand, it is known that $p_c = 1/2$.

The quantities that we consider are {\it universal}, i.e\., independent of the type
of (isotropic) percolation and the lattice structure, as long as one remains
at the percolation point $p_c$. In addition, they are  believed to have conformal invariance properties.
In particular, the crossing probabilities computed for two geometries which can be mapped
onto one another by a conformal map  should be the same. 
The universality and conformal invariance are not actually rigorously proven in the general case.  
However Smirnov \cite{Sm} has recently proven that site percolation on the triangular lattice
is conformally invariant in the scaling limit.  This work also derives Cardy's formula and confirms 
the conjectured connection of SLE and percolation (see below).  The universality (and conformal 
invariance for other percolation and lattice types) are supported by extensive numerical work and much other evidence.

At $p_c$, the probability of a configuration including a cluster spanning the lattice from left to right is
the {\it horizontal crossing probability} $\Pi_h$. The probability of a configuration  including a cluster 
connecting all four sides is the  {\it horizontal-vertical crossing probability} $\Pi_{hv}$. These  quantities
depend only on the aspect ratio $r$ because of conformal  invariance.  By the duality argument above, the 
horizontal  probability, as a function of $r$, must satisfy 
  $$  \Pi_h(r) + \Pi_h(1/r) = 1, \tag \dual  $$
while by symmetry, the horizontal-vertical probability must satisfy
  $$  \Pi_{hv}(r) \= \Pi_{hv}(1/r). \tag \s  $$

The conformal approach to the horizontal crossing proceeds by considering the $Q$-state Potts model,
expressing the crossing  as a difference of partition functions with certain non-uniform boundary conditions,
identifying the corresponding boundary CFT operators, and taking the limit $Q \to 1$.  Since the changes in boundary
conditions are implemented by boundary operators, one finds that the crossing probability is given by a four-point
  boundary operator correlation function. The horizontal-vertical crossing is obtained by a similar procedure;
the main difference is that the boundary operator is not the same as for the horizontal case.
   These derivations are fully described in the original calculations ( cf.~ \cite{Ca2}, \cite{W} or \cite{Kl}), 
and not particularly  germane to our purpose here, so we omit further details.

Instead of $r$, we can use an alternative parameter for the rectangles which is suggested by
the conformal invariance property.  If we choose a one-to-one conformal map from the rectangle onto
the unit disk (or upper half-plane; it doesn't matter), then the only conformal invariant of the
geometry is the cross-ratio $\l$ of the four points to which the corners of the rectangle are mapped. One then 
finds, by appropriate conformal manipulations, that the correlation function for horizontal crossing satisfies a
Riemann differential equation with the two solutions $F(\l) = 1$ and $F(\l) = \l^{1/3}{}_2 F_1(1/3,2/3;4/3;\l)$.
One can pick the correct linear combination by imposing the physical constraints that $F \to 0$ as $\l \to 0$ 
($r\to\infty$) and $F\to1$ as $\l\to1$ ($r\to0$). The result is {\it Cardy's formula} \cite{Ca2}
   $$ \Pi_h(r) = \frac{2\pi\sqrt3}{\Gamma (1/3)^3} \; \l^{1/3} \; {}_2 F_1(1/3,2/3;4/3;\l)\,. \tag \pih $$
The cross-ratio $\l$ is given explicitly as a function of $r$ in  Eq.~(\ltau) below. The hypergeometric function
appearing in Eq.~(\pih) is very special, since its parameters $a$, $b$, $c$ satisfy $c-a=1$, so that it reduces 
simply to the integral of an algebraic function:
   $$ \Pi_h (r) = \frac{2 \pi}{\sqrt{3} \Gamma (1/3)^3} \int_0^{\l}(t(1-t))^{-2/3}dt \,.\tag \pihint $$
This is a reflection of the fact that the hypergeometric differential equation satisfied by $\Pi_h$ factors as
  $$ \frac d{d\l}(\l(1-\l))^{2/3}\frac d{d\l}F \=0. \tag \de0 $$

For the horizontal-vertical probability $\Pi_{hv}$, the corresponding CFT analysis leads to the fifth-order differential equation
  $$ \frac{d^3}{d\l^3}(\l (1-\l))^{4/3} \frac d{d\l}(\l(1-\l))^{2/3}\frac d{d\l}F \=0 \tag \deI $$
for the function $F$ defined by  $\Pi_{hv}(r)=F(\l)$. The physical requirements of the problem are that
$\Pi_{hv}(r)$ be bounded, satisfy the symmetry condition of Eq.~(\s), which in terms of $\l$ translates into
  $$F(\l)  =F(1-\l)\,,$$
and satisfy the asymptotic condition
  $$ \lim_{r\to\infty} \frac{\Pi_{h{\bar v}}(r)}{\Pi_h (r)} = 0 \,, $$
where
  $$ \Pi_{h{\bar v}}(r)= \Pi_h(r) - \Pi_{hv}(r)  \tag \hvb $$
is the probability of there being a horizontal but no vertical crossing.
Applying these conditions to the differential equation (\deI) gives the explicit expression due to Watts \cite{W}
  $$ \Pi_{h{\bar v}}(r) = \frac{\sqrt{3}}{2 \pi} \, \l \; {}_3F_2(1,1,4/3;2,5/3;\l)\,, \tag \pihvb $$
where ${}_3F_2$ is a generalized hypergeometric function. Again, because of the special form of
the parameters, this has a simple expression as an integral:
  $$ \Pi_{h{\bar v}}(r) = \frac 1{\sqrt3\pi}\int_0^{\l}(t(1-t))^{-2/3}\int_0^{t}(u(1-u))^{-1/3}du\,dt\,. \tag\pihvbint $$

The differential equation (\deI) can also be written
  $$ \biggl(\frac{d}{d\l} \l^{-1}(1-\l)^3 \frac{d}{d\l}\l^2 \biggr)\biggl(\frac{d}{d\l}(\l(1-\l))^{1/3}
  \frac d{d\l} (\l(1-\l))^{2/3}\frac{d}{d\l} \biggr)\,F\=0\,. \tag \deII $$
This form is of interest since 1, $\Pi_h$, and $\Pi_{h\bar v}$ span the solutions of the equation formed by 
letting the rightmost factor act on $F$ alone, i.e\.
  $$ \frac{d}{d\l}(\l(1-\l))^{1/3} \frac{d}{d\l} (\l(1-\l))^{2/3} \frac{d}{d \l}F = 0\,.\tag \deIII $$
Note that the full set of solutions of Eq.(\deI) (or (\deII)) is spanned by adding the functions $\log\l$ and $\log(1-\l)$
to the three just mentioned.  Thus this problem is an example of logarithmic CFT (\cite{Gur}, \cite{Fl}, \cite{Gab}).  This behavior is to be expected on
general grounds (\cite{Ca3}), but has apparently not yet been explicitly exhibited for percolation crossing probabilities.  However, \cite{GL} calculates a closely related quantity.

In what follows, it is convenient to consider the $r$-derivatives $\Pi_h'(r)$ and $\Pi_{h{\bar v}}'(r)$
of $\Pi_h$ and $\Pi_{h{\bar v}}$. Note that these quantities are interpretable
physically as probability densities: for instance, $\Pi_h'(r)\,dr$ is the probability that the rightmost
point of any cluster attached to the left vertical side of an infinitely wide rectangle of unit height
lies between $r$ and $r+dr$ \cite{Zi1}. Note that Eq.~(\deIII) reduces to second order when considered
as a differential equation for the derivative.

\subheading{3. The Crossing Probabilities in Terms of the Aspect Ratio}

In order to set the stage for an investigation of their modular properties,
we next proceed to express $\Pi_h'$ and $\Pi_{h{\bar v}}'$ on the rectangle as
functions of the aspect ratio $r$, using the classical result for the cross-ratio, namely $\l=\l(ir)$ where
$\l(\tau)$ is the classical modular function (``Hauptmodul") for the subgroup $\Gamma(2)$ of $PSL(2,\Z)$\,.
(All needed properties of modular functions and modular forms will be reviewed in Section~4.)
This function can be given by many formulas, e.g.
$$ \l(\tau)\=16\frac{\eta(\tau/2)^8\eta(2\tau)^{16}}{\eta(\tau)^{24}}\=1- \frac{\eta(\tau/2)^{16}\eta(2\tau)^8}{\eta(\tau)^{24}}
\= \biggl(\frac{\vartheta_2(\tau)}{\vartheta_3(\tau)}\biggr)^4 \,,\tag\ltau $$
where $\eta(\tau)$, $\vartheta_2({\tau})$ and $\vartheta_3({\tau})$
are the classical modular forms of weight 1/2 (Dedekind eta function and Jacobi theta functions) defined by
$$\aligned \eta(\tau)&\=q^{1/24}\prod^{\infty}_{n=1}(1- q^n)\=\sum_{n\in\Z}(-1)^nq^{(6n+1)^2/24}\=q^{1/24}(1-q-q^2+q^5+\dots)\,,\\
\vartheta_2({\tau})&\= \sum_{n\in\Z}q^{(n+\frac12)^2/2}\=2\frac{\eta(2\tau)^2}{\eta(\tau)}\=2q^{1/8}\bigl(1+q+q^3+\dots)\,,\\
\vartheta_3({\tau})&\=\sum_{n\in\Z}q^{n^2/2}\=\frac{\eta(\tau)^5}{\eta(\tau/2)^2\eta(2\tau)^2}
  \=1+2\hat q+2{\hat q}^4+2{\hat q}^9+\dots \endaligned \tag\th $$
with $q = e^{2 \pi i \tau}$ and $\hat q:=e^{\pi i\tau}=\sqrt q$. Note that $\l=16{\hat q}-128 {\hat q}^2+704 {\hat q}^3 +\dots$
tends to 0 like $\hat q$ and that $\l$ is a power series in $\hat q$, not $q$.  (  The appearance of $\hat q$ is typical for
conformal field theory on the rectangle.)  The derivative of $\l(\tau)$ is given by
$$ \frac1{2\pi i}\l'(\tau)=8 \frac{\eta(\tau/2)^{16} \eta(2 \tau)^{16}}{\eta(\tau)^{28}}
    =8{\hat q}-128{\hat q}^2+1056{\hat q}^3+\dots\,. \tag \der $$

Now, since $\frac d{d \l}=\frac 1{\l'(\tau)} \frac d{d \tau}$, Eq.~(\deIII) can be rewritten
  $$ \frac{d}{d\tau}\frac{(\l(1-\l))^{1/3}}{\l'(\tau)} \frac{d}{d\tau}
\frac{(\l(1-\l))^{2/3}}{\l'(\tau)} \frac d{d\tau}F(\l(\tau)) = 0 $$
or, in view of  Eq.~ (\der), as the differential equation
$$ \frac{d}{d\tau}\frac{\eta(\tau)^{12}}{\eta(\tau/2)^8\eta(2\tau)^8}\frac d{d\tau}\frac1{\eta(\tau)^4}f(\tau)=0,\tag\ab$$
for the function $f(\tau):=\frac d{d \tau} F(\l(\tau))$.  From this we can immediately write down two linearly independent solutions
$$\aligned f_1(\tau) &= \eta(\tau)^4 \,,\\ 
  f_2(\tau)&=-\frac{2\pi i}3\,\eta(\tau)^4\int_\tau^\infty\frac{\eta(z/2)^8\eta(2z)^8}{\eta(z)^{12}}\,dz\,.\endaligned\tag\fIIint $$
(The factor $-2\pi i/3$, of course, is just for convenience.) 
The function $f_1$ is a modular form of weight~2 (cf.~Section 4). The function $f_2$ can be decomposed as 
  $$ f_2(\tau) = \frac1{16}\,\vartheta_2(\tau)^4\,-\,16f_W(\tau)\,,\tag \ffI$$
where $\vartheta_2(\tau)^4$, the fourth power of the theta function in (\th), is an odd function of $\hq$ and
$$ f_W(\tau)=\frac15\,q+\frac{16}{55}\,q^2+\frac{364}{935}\,q^3+\frac{13568}{21505}\,q^4
   +\frac{91614}{124729}\,q^5+\cdots\;.\tag\ffw$$
an even function of $\hq$. The function $\vartheta_2^4$ is again a modular form of weight~2, but $f_W$ is a new type
of modular object whose  transformation  properties under the modular group will be discussed in Section~7.

The $r$-derivatives of $\Pi_h$ and $\Pi_{h \bar v}$ can now be written in terms of $f_1(\tau)$ and $f_2(\tau)$ as
$$ \Pi_h'(r) = - \frac{2^{7/3}\pi^2}{\sqrt3\,\Gamma(1/3)^3}\,f_1(ir)\,,\qquad
\Pi_{h{\bar v}}'(r) = -8\sqrt3\,f_2(ir)\,.\tag{\pif}$$
(See  also \cite{Kl}.) The functions $\Pi_h$ and $\Pi_{h \bar v}$ themselves are then given by
$$ \Pi_h(r) = \frac{2^{7/3}\pi^2}{\sqrt3 \Gamma(1/3)^3} \int_r^\infty f_1(it)\,dt\,,\qquad
     \Pi_{h \bar v}(r) = 8\sqrt3\,\int_r^\infty f_2(it)\,dt\;.\tag \php  $$

Finally, we consider the recent generalization of $\Pi_h$ via SLE \cite{LSW1}.  This is a rigorous theory of 
stochastic conformal maps, driven by a Brownian process of speed $\k$, $B(\k t)$, which has been
used to calculate the Brownian intersection exponents.  
For $\k\ge0$ the hull of the process is generated by a path for all $t>0$ (\cite{RS}, \cite{LSW2}). For $0\le\k\le4$ the path is simple, while for $\k\ge8$ it is space filling. We will see in Theorem~2 below 
that the limits $\k=4$ and $\k=8$ arise from modular considerations as well.  
The corresponding horizontal crossing probability is given by a generalization of Cardy's formula,
$$F(\l;\k) = \frac{\Gamma(2-8/\k)}{\Gamma(1-4/\k)\Gamma(2-4/\k)}\,
   \l^{1-4/\k}\,{}_2 F_1\bigl(1-\frac{4}{\k},\frac{4}{\k};2-\frac{4}{\k};\l\bigr)\,.\tag\pihk $$
It is easy to show that $F(\l;\k)$ satisfies the same duality condition (Eq.~(\dual)) as Cardy's formula, and reduces
to it when $\k = 6$. 
(\cite{Sm} proves that $\k=6$ corresponds to percolation, as conjectured by \cite{LSW1}).
Further, the hypergeometric functions involved again satisfy $c-a=1$, so that one has
$$ F(\l;\k) = \frac{\Gamma(2-8/\k)}{\Gamma(1-4/\k)^2}
\int_0^{\l}\bigl(t(1-t)\bigr)^{-4/{\k}}dt \, . \tag \pihkint $$
(This also makes clear where the normalizing constant in (\pihk) comes from, since Eq.~(\pihkint) and the beta integral give
$F(1;\k)=1$.)

 There are apparently no SLE results for the horizontal-vertical crossing.  Our theorem in Section~5 gives a
candidate solution, at least up to one undetermined parameter, and includes the percolation case.

By the same arguments as in the special case $\k=6$, we can now write 
$$ \frac d{d \tau}F(\l(\tau);\k)\=\,2^{4\a}\pi i\,\frac{\Gamma(2\a)}{\Gamma(\a)^2}\,\
\frac{\eta(\tau)^{20-48 \a}}{(\eta(\tau/2) \eta(2\tau))^{8-24\a}}\,, \tag \pihpr $$
where we have set $\a=1-4/\k$ for convenience. (The rhs of (\pihpr) has a  $\hat q$-expansion beginning
with a constant times $\hq^\a$.) The modular properties of this function will be discussed in Section~5.
Integrating (\pihpr), we find the formula
$$ \Pi_h(r;\a)=2^{4\a}\pi\,\frac{\Gamma(2\a)}{\Gamma(\a)^2}\,\int_r^{\infty}\frac{\eta(it)^{20-48\a}}
{\bigl(\eta(it/2)\eta(2it)\bigr)^{8-24\a}}dt \tag \pihra $$
for the generalization $\Pi_h(r;\a):=F(\l(ir),4/(1-\a))$ of Cardy's $\Pi_h(r)=\Pi_h(r;1/3)$.

\subheading{4. Review of Modular Forms}

Let $\G_1=SL_2(\Z)$ be the group of $2\times2$ integral unimodular matrices, acting on the upper half-plane
$\H=\{\tau\in\C\mid\Im(\tau)>0\}$ by $\tau\mapsto\g(\tau)=\frac{a\tau+b}{c\tau +d}$ for $\g=\abcd\in\G_1$.
A {\it modular form} of {\it weight} $k\in\Z$ on $\G_1$ is a holomorphic function $f:\,\H\to\C$ which satisfies
$$ f\bigl(\g(\tau)\bigr) = (c\tau+d)^k\,f(\tau)\qquad\bigl(\tau\in\H,\;\g=\abcd\in\G_1\bigr) \tag \modx$$
as well as a suitable growth condition at infinity (specifically, $|f(\tau)|\le C(y^A+y^{-A})$ for some constants 
$C,\,A>0$, where $y=\Im(\tau)$).  A {\it modular function} on $\G_1$ is a meromorphic function satisfying (\modx) 
with $k=0$ (i.e., simply invariant under the action of $\G_1$ on $\H$) and a weaker growth
condition at infinity, specified below.  Every modular function can be written (in infinitely many ways) as a 
quotient of two modular forms of the same weight.  One can also consider modular forms and functions on subgroups 
$\G\subset\G_1$ of finite index, where (\modx) is required only for matrices $\g\in\G$. Other generalizations include 
allowing a {\it character} by including a factor $v(\g)$ on the rhs of Eq.~(\modx), where $|v(\g)|=1$ for all $\g\in\G$, 
or allowing $k$ to be a half-integer.  In the case of the full modular group $\G_1$, the collection of equations (\modx) 
can be replaced by the two equations $f(\tau+1)=f(\tau)$ and $f(-1/\tau)=\tau^k f(\tau)$, since $\G_1$ is generated by 
the two matrices $T = \bigl(\smallmatrix 1&1\\0&1\endsmallmatrix\bigr)$ and $S=\bigl(\smallmatrix 0&-1\\1&0\endsmallmatrix\bigr)$.
These matrices (or rather, the automorphisms of $\H$ which they represent) satisfy the relation $S^2=(ST)^3=1$. A second 
group which will play an important role for us is the {\it theta group} $\G_\theta$ consisting of matrices in $\G_1$ congruent
to $1$ or $S$ modulo 2.  It is generated by the two matrices $S$ and $T^2$.

We can restate the modular invariance property (\modx) conveniently as $f|_k\gamma=f$, with the action ``$|_k\g$" of
$\g=\abcd\in\G_1$ on functions on $\H$ defined by $(f|_k\gamma)(\tau)=(c\tau+d)^{-k}f(\gamma(\tau))$.  This shorthand
notation will be convenient in what follows.

It follows from Eq.~(\modx) that modular forms of a given weight are a
vector space over $\C$, denoted $M_k(\Gamma_1)$ for forms defined over the full modular group
$\Gamma_1$, or more generally $M_k(\Gamma,v)$. The dimension of this space  for simple groups $\G$ and small values 
of $k$ is very small,  a fact which leads to many non-trivial identities, a few of which we exploit below. Modular forms 
have power series expansions in non-negative powers of $q=e^{2\pi i\tau}$. Modular forms $f$ over $\Gamma_1$ whose 
$q$-expansion has no constant term, so that $f \to 0$ as $\tau \to \infty$, are called {\it cusp forms}. In general, 
cusp forms of weight $k$ on subgroups $\G\subseteq\G_1$ are defined by requiring that  $|f(\tau)| \le C\,\Im(\tau)^{k/2}$  
for some $C>0$ and all $\tau\in\H$. For each weight $k$ they form a subspace $S_k(\G)\subseteq M_k(\G)$. As examples,
the spaces $M_2(\G_\theta)$ and $S_{12}(\G_1)$ are both one-dimensional, with generators $\vartheta_3(\tau)^4$ and
$\eta(\tau)^{24}$, respectively, while $\l'(\tau)$ satisfies $\l'|_2T^2=\l'$, $\l'|_2S=-\l'$ and hence is a cusp form
of weight~2 on $\G_\theta$ with a non-trivial character.

An important theorem for a modular form $f \in M_k(\Gamma_1,v)$ expresses a kind of ``sum rule"
on the total number of zeros.  It may be obtained by integrating the logarithmic derivative of f around the boundary 
of a fundamental domain $F=\H/\Gamma_1$ (for more details see \cite{Ko}). Let $\nu_P(f)$ denote the order of the zero 
of $f(\tau)$ at the point $P\in \H$, while $\nu_\infty(f)$ denotes the exponent of the leading term in the $q$-expansion 
of $f$, i.e.~$\nu_\infty(f)=\a$ if $f(\tau)$ has a Fourier expansion of the form $\sum_{n=0}^\infty a_nq^{n+\a}$ with
$a_0\ne0$. Then
$$ \nu_\infty(f)+\frac12\nu_i(f)+\frac13\nu_{\rho}(f)+\sum_{P \in\H/\Gamma_1,\;P\ne i,\rho}\nu_P(f)=\frac{k}{12} \tag \nozs $$
Note that $\nu_i(f), \nu_{\rho}(f), \nu_P(f)$ are non-negative integers, while $\nu_\infty(f)$ is integral if $k$ is even
and $v$ is trivial but can be a rational or even real number in general. (If $\nu_\infty(f)=\a$ then $v(T)=e^{2\pi i\a}$.) 
The points $i$ and $\rho=\frac12(1+i\sqrt3)$ are fixed by the elements $S$ and $ST$ of order $2$ and $3$, respectively,
and thus are differently weighted.  A similar formula applies if $f\in M_k(\Gamma,v)$ for any subroup $\G$ of $\G_1$,
but the rhs is multiplied by the index of $\Gamma$ in $\Gamma_1$ and a different set of points and weightings appear on the lhs.
In particular, for $f\in M_k(\Gamma_\theta,v)$, Eq.~(\nozs) becomes
$$ \nu_\infty(f)+\nu_1(f)+\frac12\nu_i(f)+\sum_{P \in \H/\Gamma_{\theta},\;P\ne i}\nu_P(f)=\frac{k}{4}\,, \tag\nogzs $$
where $\nu_\infty(f)$ is now defined by $\nu_\infty(f)=\a$ if $f=\sum_{n=0}^\infty a_n\hq^{n+\a}$ with $a_0\ne0$,
since the local parameter at $\infty$ is $\hat q$ not $q$,  and $\nu_1(f)$ is similarly defined as the leading power of $q$
in $f|_k T^{-1} S$.  Again $\nu_i(f)$ and $\nu_P(f)$ are (non-negative) integers,
while $\nu_1(f)$ and $\nu_\infty(f)$ can be arbitrary (non-negative) real numbers.

One can also look at meromorphic modular forms.  Now $\nu_P(f)$ both at ``finite" points $P\in\H$ and at cusps
like $P=\infty$ or $1$ are still required to be finite but may be negative. (For $k=0$ this is the ``weaker growth 
condition at infinity" for modular functions  mentioned at the beginning of the section.) The sum rules given above 
still hold in this more general context.  The presence of poles is reflected in the growth of the coefficients of the 
$q$-expansion as follows: these coefficients have polynomial growth in the case of holomorphic modular forms, grow 
exponentially in the square root of the index if the function is holomorphic in $\H$ but has poles at the cusps (for 
instance, for the modular function $\l(\tau)$), and have exponential growth if the function has poles at finite points.

\subheading{5. Modular Properties of the Crossing Probabilities}

The purpose of this section is to show that the functions found by Cardy \cite{Ca2}, Watts \cite{W}, and Lawler et al
\cite{LSW1} as the solutions of crossing probability problems are  characterized by certain very simple
mathematical properties. This naturally raises the question (to which we do not know the answer) whether these
properties can be seen by some {\it a priori} arguments for the crossing probabilities,
in which case our theorems would provide very simple new derivations of the
results of these authors.  Our results also include a possible generalization of Watts's
formula for the ``horizontal-vertical" crossing probability to the SLE processes.

We will call a function $\Pi$ on the positive real axis a {\it conformal block} if it is expressible as a (real) 
power of $\hat q = e^{-\pi r}$  times a power series in $\hat q$. More precisely, if
$$ \Pi(r) = \sum_{n=0}^{\infty}a_n\hq^{n+\a} \tag \cblock $$
with $\a \in \R$ and $a_0 \neq 0$, we call $\Pi(r)$ a conformal block of dimension $\a$.  This type of 
function is ubiquitous in CFT on a rectangle, appearing in partition functions and correlation functions as well.  
We will call $\Pi(r)$ an {\it even conformal block} if $a_n = 0$ for $n$ odd, so that $\Pi(r)$ equals $\hq^\a$  
times a power series in $q=\hq^2$.  Cardy's crossing probability $\Pi_h(r)$  satisfies this stronger condition.  
Notice that the convergence of the series in (\cblock) for all $r>0$ implies that $a_n=\text O(c^n)$ for  any $c>1$ 
and hence the corresponding function  in the upper half-plane (defined by the same series but with $\hq$ replaced 
by $e^{\pi i\tau}$) is holomorphic and is an eigenfunction of the operator $T^2:\tau\mapsto\tau+2$ or, in the 
case of an even conformal block, even of $T:\tau\mapsto\tau+1$.

Note that the nomenclature ``conformal" is only suggestive at this point, since this definition
does not imply that  $\Pi(r)$ is related to any CFT model.

We are now in a position to state
\proclaim{Theorem 1}
Let $\Pi(r)$ be any function on the positive real axis such that
\roster\item"(i)" $\Pi(r)$   is an even conformal block with dimension $\a > 0$;
\item"(ii)" $\Pi(1/r)=1-\Pi(r)$.
\endroster
Then $\a = 1/3$ and $\Pi(r)$ is Cardy's function. \endproclaim
\demo{Proof} Define $\{a_n\}$ by (\cblock), and define $P(\tau)$ (for $\tau \in  \H$) by the same expression 
as in (\cblock) but with $\hq$ interpreted as $e^{\pi i\tau}$ rather than $e^{-\pi r}$, so  that $P(ir)=\Pi(r)$
for $r>0$. It follows that $P(\tau)$ is analytic in the whole upper half-plane.  By property (ii) and analytic 
continuation, we have $P(-1/\tau)=1-P(\tau)$, while the fact that $a_n=0$ for $n$ odd gives $P(\tau+1)=A P(\tau)$, 
where $A=e^{\pi i \a}$.  Hence $f(\tau):=P'(\tau)$ is holomorphic in $\H$ and satisfies $f|_2S=-f$ and $f|_2T=Af$.  
Also, $f$ is small at infinity because of the assumption $\a>0$. At this point there are two ways to complete the argument: 
\roster\item"{\bf A:}"  From $f|S = -f$,  $f|T = Af$  we deduce  $f|(ST)^3 = -A^3 f$  and hence, since
$(ST)^3=1$, that $A^3=-1$, i.e., $\a=m/3$ where $m$ is an odd integer. It follows that $f^6 \in S_{12}(\Gamma_1)$. 
But this space has dimension~1 and is spanned by $\eta^{24}$, as mentioned in Section~4, so $f=C\eta^4$ for some  $C\ne0$.  
The integration constant required follows from the condition~(ii).

\item"{\bf B:}" From the modular properties mentioned, $f\in M_2(\Gamma_1,v)$ for some character $v$. The rhs of Eq.~(\nozs)
is therefore $1/6$. Since each term on the lhs is non-negative and $\nu_i$, $\nu_\rho$ and $\nu_P$ are integers, the only
possibility is  $\nu_i = \nu_\rho = \nu_P = 0$ and $\nu_{\infty} =\a/2=1/6$.   It follows that $A^6=1$ and that the 
quotient of $f^6$ by $\eta^{24}$ is holomorphic, bounded, and $\G_1$-invariant, and therefore constant. 
\endroster \enddemo

Consider the percolation crossing problem. It is interesting that if  one assumes, following Theorem 1, that the horizontal crossing 
probability is given by a {\it single} conformal block, its evenness  follows from the physics, since the boundary conditions used 
in the conformal analysis are the same on the two horizontal sides (see \cite{Ca4} for further discussion of this point).

Our next theorem generalizes Theorem 1 by dropping the assumptions of evenness and positivity of $\a$, but one has
to add a growth condition on the coefficients of $\Pi(r)$ which was automatic in the even case. 

\proclaim{Theorem 2} Let $\Pi_1(r)$ be any function on the positive real axis such that
\roster\item"(i$'$)"  $\Pi_1(r)$ is a conformal block of dimension $\a\in\R$ with coefficients $a_n$ of polynomial growth;
\item"(ii)" $\Pi_1(1/r)=1-\Pi_1(r)$.
\endroster
Then $0<\a\le 1/2$ and $\Pi_1(r)=\Pi_h(r;\a)$, the generalized Cardy's function of Eq.~$(\pihra)$. \endproclaim
\demo{Proof} We argue as in Method B above. First we define $P_1(\tau)$  similarly to $P(\tau)$.  Note that it is a 
conformal block but not necessarily even. Its derivative $f_1(\tau):=P_1'(\tau)$ is holomorphic and of polynomial growth 
because of the polynomial growth assumption on the $a_n$, and satisfies $f_1|_2S=-f_1$ and $f_1|T^2=Af_1$, with $A=e^{2\pi i\a}$. 
Since $S$ and $T^2$ generate $\G_\theta$, it follows that $f_1 \in M_2(\Gamma_\theta,v)$ for some $v$.  We now apply Eq.~(\nogzs).  
The rhs is 1/2. The assumption of polynomial growth implies that $\nu_\infty(f)$ and $\nu_1(f)$ are non-negative
real numbers, while $\nu_i(f)$ and $\nu_P(f)$ are non-negative integers.  On the other hand, $\nu_\infty(f)$
equals $\a$ if $\a\ne0$ and is $\ge1$ if $\a=0$.  Eq.~(\nogzs) therefore implies that $0<\a\le\frac12$,
$\nu_\infty(f)=\a$, $\nu_1(f)=\frac12-\a$ and all other $\nu_P$ vanish.  Since the rhs of (\pihpr) has the
same properties, these two functions must be proportional (their ratio is a function on a compact Riemann surface with
no zeros or poles, hence constant); and then since $\Pi_1(r)$ vanishes at infinity it must
be proportional (and hence, by (ii), equal) to $\Pi_h(r;\a)$.
\enddemo

We make a few comments about the interpretation of the lower and upper bounds $\a=0$ and 
$\a=1/2$ in Theorem 2 from the modular and from the physics point of view. In the proof of
the theorem, both bounds arose from the requirment that the modular form of weight 2 given by
(\pihpr) be ``holomorphic at the cusps," i.e., that the two numbers $\nu_\infty(f)$ and $\nu_1(f)$ in
(\nogzs) should both be non-negative.  If $\a<0$, then the leading power in the $\hq$-expansion of 
this function is negative and the integral in (\pihra) diverges.  If $\a>1/2$, then the integral
converges and  the function $\Pi_h(r;\a)$ still satisfies the functional equations in (i) and
(ii), but it no longer satisfies the growth assumption: now $\nu_1(P_1')<0\,$ and the $a_n$ grow 
exponentially in $\sqrt n$ as explained at the end of~Section~4. (In Theorem~1 we did not have to 
explicitly make the assumption of polynomial growth because in this case there was only one
cusp---compare equations (\nozs) and (\nogzs)---so that the assumption $\a>0$ already implied the 
holomorphy of the function $f(\tau)=P'(\tau)$.)  From the physics point of view the assumption
$\a>0$ is natural since the probability $\Pi(r)$ has to go to zero for large~$r$, but the physical 
meaning of the polynomial growth condition is not obvious.  However, both critical values
have a physical meaning in terms of the SLE processes, as already mentioned in~Section~3: the lower
limit $\a=0$ corresponds to the value $\k = 4$ above which the hull of the process is no longer generated by
a simple path, while the upper limit $\a=1/2$ corresponds to the value $\k=8$ above which 
the path becomes space filling.  At intermediate values, namely $4<\k<8$, the path is self-intersecting.

Our final theorem in this section reproduces and generalizes Watts's formula for the percolation
crossing probability $\Pi_{hv}$.

\proclaim{Theorem 3} Let $\a$ and $\Pi_1(r)$ be as in Theorem $2$ and $\Pi_2(r)$ be a second function satisfying
\roster\item"(iii)" $\Pi_{2}(r)={e}^{-\pi\b r}\sum_{n=0}^{\infty}b_n{e}^{-\pi nr}$ for some $\b\in\R$,
   with $\{b_n\}$ of polynomial growth;
\item"(iv)" $\Pi_-(1/r)=\Pi_-(r)$, where $\Pi_-:= \Pi_1-\Pi_2$. \endroster
Then \nl a $0<\b\le1$, $\b\ne\a\,$.  
 \nl b The function $\Pi_-(r)$ is given by the formula
$$ \Pi_-(r)\=\,C(\a,\b)\,\int_r^{\infty}\frac{\eta(it)^{20-48\a}}{(\eta(it/2) \eta(2it))^{8-24\a}}
   \int_1^t\frac{\eta(iu)^{20-48(\b-\a)}} {(\eta(iu/2) \eta(2iu))^{8-24(\b-\a)}}\,du\,dt\;,\tag \newi $$
with  $$C(\a,\b)=2^{4\b+1}\pi^2\,\frac{\G(2\a)\G(2\b-2\a)}{\G(\a)^2\G(\b-\a)^2}\;. \tag\const $$ 
  \nl c If also $\Pi_2(r)$ and $\Pi_-(r)$ are positive for all $r>0$, then $\b>\a$ and 
$$ \Pi_2(r)\=\,C(\a,\b)\,\int_r^{\infty}\frac{\eta(it)^{20-48\a}}{(\eta(it/2) \eta(2it))^{8-24\a}}
\int_t^\infty\frac{\eta(iu)^{20-48(\b-\a)}} {(\eta(iu/2) \eta(2iu))^{8-24(\b-\a)}}\,du\,dt\;.\tag \thiii $$
\endproclaim
The functions $\Pi_1$, $\Pi_2$ and $\Pi_-$ are intended to be the generalizations of $\Pi_h$, $\Pi_{h\bar v}$ and $\Pi_{hv}$, 
respectively, and (\thiii) agrees with the second formula in (\php) in the case $\a=1/3, \b=1$.
\demo{Proof} The argument again follows the proof of Theorem 1, Method B. First we define $P_1(\tau)$ as in 
Theorem 2, and $P_2$ and $P_-$ analogously.  Both these functions and their first derivatives $f_1$, 
$f_2$ and $f_-$ are holomorphic and of polynomial growth in $\H$. The modular transformation equations 
of $f_1$ were given in the proof of Theorem 2, while  $f_-|_2 S=f_-$ and $f_2|T^2=Bf_2$, with
$B=e^{2\pi i \b}$. Thus the function $v=f_2/f_1$ satisfies  $v|_0 S=2-v$ and $v|T^2=(B/A)v$, so the function
$g:=v'f_1=f_-(f_1'/f_1)-f_-'$ satisfies $g|_4S=g$, and $g|T^2=Bg$. But $g$ is also holomorphic and of
polynomial growth (because $f_1$, the modular form given in eq.~(\pihpr), has no zeros in $\H$ and only 
exponential growth at infinity), so $g\in M_4(\Gamma_\theta,v)$ for some character $v$.  We now apply 
Eq.~(\nogzs) to $g$. The rhs is 1, so, since all terms  on the lhs are non-negative and all except 
$\nu_\infty=\b$ and $\nu_1$ are integral (the equation $g|_4S=g$ implies that $\nu_i(g)$ is even!), 
we must have $0\le\b\le1$, $\nu_1=1-\b$ and $\nu_P(g)=0$ for all $P\in\H$.  The case $\b=0$ can
be excluded since then the expansion of the function $P_2(\tau)$ would begin with a constant term and its 
derivative $f_2$ would have order $\b+1/2=1/2$ rather than $\b$ at infinity, and the orders of $v$ and $g$ 
at infinity would become  $1/2-\a$ and $1/2$ respectively, leading to a contradiction with~(\nogzs).  
Hence $0<\b\le1$. The fact that $g$ is modular of weight 4 on $\G_\theta$ and has the given orders 
of vanishing at all (finite and infinite) points now fixes it uniquely up to a constant: we must have  
  $$ g(\tau)\;\=\;C_1\,\frac{\eta(\tau)^{40-48\b}} {(\eta(\tau/2) \eta(2\tau))^{16-24\b}} $$
for some non-zero complex number $C_1$ (again, because the ratio of the functions on the left and the
right is a meromorphic function with no zeros and poles and hence constant) and therefore
  $$ v'(\tau)\=\frac g{f_1}\;=\;C_2\,\frac{\eta(\tau)^{20-48(\b-\a)}} {(\eta(\tau/2)
      \eta(2\tau))^{8-24(\b-\a)}} \tag\vder$$
for some non-zero complex number $C_2$. To complete the argument, we must integrate Eq.~(\vder), multiply the 
result by $f_1$ to obtain $f_2$, integrate again, and then adjust the constants, if possible, so that
all the conditions of the theorem are satisfied.  

We first note that the transformation equation $v(\tau)+v(-1/\tau)=2$
implies that $v(i)=1$, so that (\vder) integrates to
  $$ v(\tau)\= 1 + C_2\,\int_i^\tau \frac{\eta(\tau')^{20-48(\b-\a)}} {(\eta(\tau'/2)
      \eta(2\tau'))^{8-24(\b-\a)}}\,d\tau'\,. \tag\vint $$ 
Since also  $ P_-'(\tau)=f_-(\tau)\=f_1(\tau)\,(1-v(\tau))$ and $\Pi_-(r)$ is small at infinity (because both
$\Pi_1(r)$ and $\Pi_2(r)$ are), this implies formula (\newi) except for the determination of the constant $C(\a,\b)$.
Next, we look at the behavior of the functions at infinity. The right-hand side of (\vder) has a
$\hq$-expansion beginning $C_2\,\hq^{\b-\a}(1+(8-24\b+24\a)\hq+\cdots)\,$.  It follows immediately that $\a\ne\b$,
completing the proof of (a) of the theorem, because if $\a$ were equal to $\b$ then we would have
$v'(\tau)=C_2+\text O(\hq)$ and hence $v(\tau)=C_2\tau+\text O(1)$, contradicting the periodicity
($T^2$-invariance) of $v$.  If $\b>\a$, then $v'(\tau)$ is exponentially small at infinity and we can 
integrate (\vder) to get 
$$ v(\tau)\= C_3 - C_2\,\int_\tau^\infty \frac{\eta(\tau')^{20-48(\b-\a)}} {(\eta(\tau'/2)
      \eta(2\tau'))^{8-24(\b-\a)}}\,d\tau'\,. \tag\vintii $$ 
instead of (\vint).  The constant $C_3$ here must be 0 because the $\hq$-expansion of $v=f_2/f_1$ should begin
with a multiple of $\hq^{\b-\a}$, not $\hq^0$.  To get the value of the constant $C_2$, we compare (\vintii) 
(with $C_3=0$) and (\vint), obtaining:
  $$ C_2\,\int_i^\infty \frac{\eta(\tau')^{20-48(\b-\a)}} {(\eta(\tau'/2)
      \eta(2\tau'))^{8-24(\b-\a)}}\,d\tau'\=-1\,.$$
Using  Eqs.~(\ltau) and (\der), we can rewrite this by setting $u=\l(\tau')$ as 
$$  C_2\,\int_0^{1/2} (u(1-u))^{\b-\a-1}\,du\=-16^{\b-\a}\pi i\,.$$
But from the invariance of $u(1-u)$ under $u\mapsto1-u$ and the standard beta integral we have
$$\int_0^{1/2} (u(1-u))^{\b-\a-1}\,du\=\h\,\int_0^1(u(1-u))^{\b-\a-1}\,du\=\h\frac{\G(\b-\a)^2}{\G(2\b-2\a)}$$
This gives the formula 
$$C_2\= -2^{4\b-4\a-1}\,\frac{\G(2\b-2\a)}{\G(\b-\a)^2}\,\pi i\,$$ 
for $\b>\a$, and since the function defined by (\vint) must have a $\hq$-expansion of the form 
$\,c_0\hq^{\b-\a}+0\cdot\hq^0 \allowmathbreak +c_1\hq^{\b-\a+1}+\cdots\,$ for all values of $\b$ (again, because of the
requirement that $v$, and not merely $v'$, be a single conformal block), it follows by analytic continuation
that the same formula is true also for $\b<\a$. We note, in any case, that under the assumption that $\Pi_-(r)$
and $\Pi_2(r)$ are both non-negative for real $r$ (which is certainly what we want in the physical situation,
since these functions are meant to represent probabilities), we have $0<\Pi_2(r)<\Pi_1(r)$ and hence automatically
$\b>\a$ (let $r\to\infty\,$!), as stated in (c).   Finally, the value of the constant $C(\a,\b)$ in (\newi) 
and (\thiii) follows from the above formula for $C_2$ together with the requirement that $\Pi_-$ and $\Pi_2$ should
add up to $\Pi_1$, whose normalizing constant was already given in Theorem~2, and analytic continuation to include 
the case $\b<\a$.  This completes the proof of the theorem.   
\enddemo

The arguments in the last part of the proof say that the formulas (\newi)--(\thiii) can be rewritten in 
terms of the variable $\l=\l(r)$ as $\Pi_-(r)=\Pi_{hv}(\l;\a,\b)$ and $\Pi_2(r)=\Pi_{h\bar v}(\l;\a,\b)$, where 
$$ \Pi_{hv}(\l;\a,\beta) \=2\,\frac{\G(2\a)\G(2\b-2\a)}{\G(\a)^2\G(\b-\a)^2}
         \int_0^{\l}(t(1-t))^{\a-1}\int_t^{1/2}(u(1-u))^{\beta-\a-1}\,du\,dt\,,$$
$$  \Pi_{h\bar v}(\l;\a,\beta) \= 2\,\frac{\G(2\a)\G(2\b-2\a)}{\G(\a)^2\G(\b-\a)^2}
         \int_0^{\l}(t(1-t))^{\a-1}\int_0^t(u(1-u))^{\beta-\a-1}\,du\,dt\,. $$
These two functions add up to the function $\Pi_h(r;\a)$ of (\pihra) (or (\pihkint) with $\k=4/(1-\a)$), and all three 
satisfy the differential equation
$$   \frac{d}{d\l}(\l(1-\l))^{1+\a-\beta} \frac{d}{d\l} (\l(1-\l))^{1-\a} \frac{d}{d \l}F = 0\,, \tag \deIIIp $$
generalizing Eq.~(\deIII). The space of solutions of this equation is spanned by 1, $\Pi_h(r;\a)$, and $\Pi_{hv}(r;\a,\beta)$.

Theorem~3 picks out for each $\a$ certain special functions $\Pi_{hv}(r;\a,\b)$ and $\Pi_{h\bar v}(r;\a,\b)$, 
related to $\Pi_h(r;\a)$ in a nice way, by relatively simple axiomatic properties. The hope, of course, as
already suggested by the notation, is that for a suitable value of $\b$ these functions really may give the 
correct horizontal-vertical crossing probability for the SLE process with $\k=4/(1-\a)$, although we have 
no real basis for this beyond its validity in the special case $\a=1/3$.  (In particular, there seems 
to be no obvious reason from the physics why $\Pi_{h\bar v}$ should be a single conformal block.)  Even assuming 
that it is true, we are still left with the problem of determining or guessing what the correct value
of $\b$ is.  For $\a=1/3$ we know that we must take $\b=1$.  If we make the simplest possible
assumption that $\b$ is given by a linear function of $\a$, and if there really is a $\b$ corresponding
to $\a$ for every $\a$ in the interval $(0,1/2)$, then we must always have $\b=1$, because this is the 
only line through $(1/3,1)$ contained in the box $(0,1/2]\times(0,1]$ permitted by Theorem~3.  
Another special feature of the value $\b=1$ is that here, and for no other value of $\b>0$, the differential
equation (\deIIIp) reduces to a purely hypergeometric one, so that we have the formula
 $$\Pi_{h{\bar v}}(r;\a,1)\=\frac{\tan\pi\a}\pi\,\frac{1-2\a}{1-\a}\,\cdot\,\l\;
     {}_3F_2(1,1,2-2\a;2,2-\a;\l)\,,\tag\pihvbnew$$
generalizing (\pihvb), with $\l=\l(r)$.  Just as in the discussion of the special case $\a=1/3$ in Section~2, the
hypergeometric equation occurring here is highly degenerate and its three fundamental solutions have the simple form
$$ 1\,,\quad\;\sum_{n=0}^\infty \frac{(n-\a)!}{n!}\,\frac{\l^{n+\a}}{n+\a}\,,\quad\;
       \sum_{n=1}^\infty\frac{(n-2\a)!}{(n-\a)!}\,\frac{\l^n}n $$
(where $x!:=\G(x+1)$), with power series whose coefficients involve only two gamma-functions rather than six
as would be the case for a generic $\,{}_3F_2\,$.

Note that SLE corresponds to the Q-state Potts models via $Q=4\, cos^2(4\pi/\k)$, for $\k\ge4$, as conjectured in \cite{RS} (see also \cite{BB} and \cite{FW}).  Thus quantities such as the horizontal-vertical crossing probability, specified in the Potts case, should also exist in SLE.

\subheading{6. Modular Properties of the Partition Function}

In this section, which is directed more at readers conversant with the basics of conformal field theory, we briefly discuss
a different situation in which modularity arguments can be used in statistical mechanics.  The partition function of any conformally
invariant system defined on an $l\times l'$ rectangle (with edges) is supposed to have a universal common factor $Z(l,l')$.
Assuming the same conformally invariant boundary condition on each edge, it is possible to
calculate this factor by use of CFT.   Up to an undetermined multiplicative real constant $C$, the result \cite{KlV}  is
  $$ Z(l,l')\;=\;C\,l^{c/4}\,\eta(\tau)^{-c/2}\,, \tag \rect $$
where $\tau := i(l'/l)$ and $c\in\R$ is the conformal central charge.  The rhs of this is a modular form (of real weight,
with character, and possibly with poles at infinity) on the full modular group $\G_1$, so one can ask whether it is possible
to reproduce $Z$ from modular considerations. This we now proceed to do, using certain assumptions based on the physics of 
the problem, thus showing that modular properties of conformal quantities defined on rectangles appear in a wider range of
problems than just crossing probabilities. The theorem below is a variant of an unpublished argument due to Cardy.

First, it is obvious from the definition of $Z$ that it satisfies (i) $Z(l,l')=Z(l',l)$. Next, because of the ``trace anomaly"
effect due to the corners of the rectangle \cite{CP}, we know that~$Z$ is homogeneous of degree $c/4$, so we can write
(ii) $Z(l,l')=l^{c/4}d(\tau)$ for some function $d(\tau)$.  Finally, if we assume that only one conformal block contributes to $Z$, 
then we have (iii) $d(\tau)=q^{-c/48}\sum_0^{\infty}a_n{q}^n$, for all $l'/l>0$, where $a_0\ne0$ and $q= e^{2\pi i\tau}$ as usual,
because the leading behavior of $Z$ as $\tau \to \infty$ follows simply from known results for the partition function on an 
infinite strip and because the fact that the boundary conditions are the same on all sides of the rectangle implies that any 
single conformal block must be even (by the arguments in \cite{Ca4}).

\proclaim{Theorem 4}
Any function $Z(l,l')$ which satisfies the conditions (i), (ii) and (iii) above must be given by eq.~$(\rect)$.
\endproclaim
\demo{Proof} The convergence of the series in (iii) for $\tau\in i\R_+$ implies its convergence for all $\tau\in\H$, so
$d(\tau)$ is a holomorphic function in $\H$.  Substituting (i) into (ii) we find that the product $h(\tau)=\eta(\tau)^{c/2}d(\tau)$
is invariant under $\tau\mapsto-1/\tau$ (first for $\tau/i$ real, and then by analytic continuation for all $\tau\in\H$),
while (iii) implies that $h(\tau)$ is also invariant under $\tau\mapsto\tau+1$ and hence under the whole modular group $\G_1$.
Since assumption (iii) also implies that $h$ is bounded at infinity, and since the group $\G_1$ has only one cusp, $h$ is
bounded on all of $\H/\G_1$ and hence constant.  

Notice that if we had weakened assumption (iii) to just ``$d(\tau)$  is a
single even conformal block, of dimension $\a$," then we could still deduce that $\a=-c/48+2n$ for some integer $n\ge0$,
and hence that $Z$ is given by (\rect) if we assumed $\a<-c/48+2$, while otherwise $d(\tau)$ would in general be the 
product of $\eta(\tau)^{-c/2}$ and a polynomial of degree $\le n$ in the modular invariant function $j(\tau)$. \enddemo

\subheading{7. Eichler Integrals and Higher Order Modular Forms}

In this section we describe in a little more detail the modular properties of the crossing probabilities studied in
this paper.  We will concentrate mostly on the functions $\Pi_h(r)$ and $\Pi_{h{\bar v}}(r)$ of the original
percolation problem as given in eq.~(20), but analogous remarks would apply also to the functions occurring in
Theorem~3 for values of~$\a$ and~$\beta$ other than~1/3 and~1.

In the theory of modular forms, the {\it Eichler integral} of a modular form $f(\tau)$ of integral weight $k\ge1$ is
a $(k-1)$st primitive of $f$, i.e., a function $\tf(\tau)$ in the upper half-plane whose $(k-1)$st derivative 
is (a multiple of) $f(\tau)$.  If $f(\tau)=\sum a_n\,\hq^{n+\a}$, then $\tf(\tau)$ can be given explicitly by
$\tf(\tau)=\int_\tau^\infty(z-\tau)^{k-2}f(z)\,dz$ or by $\tf(\tau)=\sum(n+\a)^{1-k}a_n\,\hq^{n+\a}$.  These functions
are no longer modular, but are ``nearly modular" of weight $2-k\,$: if $f$ is modular with respect to $\G$, then
$(c\tau+d)^{k-2}\tf\bigl(\g(\tau))$ is the sum of $\tf(\tau)$ and a polynomial in $\tau$ of degree $k-2$ for each
matrix $\g=\abcd\in\G$. In the special case $k=2$, $\tf(\tau)$ is (up to a constant) simply the integral of $f$
from $\tau$ to $\infty$ and transforms via $\tf(\g(\tau))=\tf(\tau)+C(\g)$ for all $\g\in\G$, where $\g\mapsto C(\g)$
is a homomorphism from $\G$ to $\C$.  In certain cases (namely, when $f$ is a Hecke eigenform with integral eigenvalues;
we do not explain the details), the image of this homomorphism is a lattice $\L\subset\C$ and the function $\tf$
gives a map from the {\it modular curve} $\H/\G$ to the {\it elliptic curve} $\C/\L$.

In the case of the Cardy function, we see from equations (\fIIint) and (\pif) or~(\php) that $\Pi_h(r)$ equals $\tf_1(ir)$, where
$\tf_1(\tau)$ is the (suitably normalized) Eichler integral associated to the weight~2 modular form $f_1(\tau)=\eta(\tau)^4$.
The function $\tf_1$ gives a modular parametrization of the elliptic curve $\C/L$, where $L$ is the lattice spanned
by~1 and $\frac12+\frac16\sqrt3$. This elliptic curve has the Weierstrass equation $Y^2=X^3+1$. The constant involving
$\G(\frac13)^3$ in front of the first integral in (\php) is essentially the reciprocal of one of the ``periods" associated
to this elliptic curve.  (Again, we omit details.)

We now turn to the second function $f_2(\tau)$ in (\fIIint), which is a less familiar type of modular object: it is not
modular, but its failure to be modular is given simply by multiples of the modular form $f_1\,$.  More precisely, we have
$$  f_2(\tau+2)\,=\,f_2(\tau)\,,\qquad \tau^{-2}\,f_2(-1/\tau)\,=\,f_2(\tau)\,-\,C\,f_1(\tau)\,,\tag{\hof}$$
where $C=2^{1/3}\pi^2/3\G(1/3)^3$.  To see this, we observe that $f_2$ is the product of $f_1$ with the Eichler
integral $\tf_3$ of the modular form $f_3(\tau)=\eta(\tau/2)^8\eta(2\tau)^8/\eta(\tau)^{12}$ of weight 2. The
function $f_3$ transforms under the generators of $\G_\theta$ by $f_3(\tau+2)=\omega^2f_3(\tau)$ and $\tau^{-2}f_3(-1/\tau)
=-f_3(\tau)$, where $\omega = e^{2\pi i/3}$, so its integral $\tf_3$ transforms by $\tf_3(\tau+2)=\omega^2\tf_3(\tau)+c_1$ 
and $\tf_3(-1/\tau)=-\tf_3(\tau)+c_2$ for some constants of integration $c_1$ and $c_2$, the first of which is easily
seen to be 0. Multiplying these equations by $f_1(\tau+2)=\omega f_1(\tau)$ and $\tau^{-2}f_1(-1/\tau)=-f_1(\tau)$, one
finds equation (\hof).  Note that the transformation properties in (\hof) are the ones which were used and generalized in
Theorem~3 (in particular, the function occurring in (32) specializes to $f_3$ for $\a=1/3$, $\b=1$).  Notice also that
the second equation in (\hof), which can be written as $f_2|S=f_2-Cf_1$, can be combined with the transformation property
$f_1|S=-f_1$ to say that the linear combination $f_2-\frac12Cf_1$ is invariant under $S$.  In terms of the original
problem, the functions $f_1$, $f_2$ and $f_2-\frac12Cf_1$ are proportional to the derivatives of $\Pi_h$, $\Pi_{h\bar v}$
and $\Pi_{hv}$, respectively, and this last property is just a restatement of (the derivative of) equation (\s). Finally,
we can combine the two equations (\hof) by saying that the vector $F=\pmatrix f_1 \\f_2 \endpmatrix$ transforms under 
$T^2$ and $S$ by
  $$ F(\tau+2)=\pmatrix\omega&0\\0&1\endpmatrix F(\tau)\,,\qquad
     \tau^{-2}\,F(-1/\tau)=\pmatrix-1&0\\-C&1\endpmatrix F(\tau)\,,\tag{\hoff}$$
i.e., it is a vector-valued modular form of weight~2 on the group $\G_\theta$.

Following this example, we define a {\it second order modular form} of weight $k$ with respect to a subgroup
$\G$ of $SL_2(\Z)$ to be a holomophic function $f(\tau)$ which satisfies  $f|_k(\g_1-1)|_k(\g_2-1)=0$ for all
$\g_1,\g_2 \in \G$, or equivalently, if $f|_k(\g-1)$ is modular of weight $k$ for all $\g\in\G$ (rather than being~0
for all $\g$ as for an ordinary modular form). More generally, an {\it $n$-th order modular form} is a function satisfying 
$f|_k(\g_1-1)\cdots|_k(\g_n-1)=0$ for all $\g_1,\ldots,\g_n\in\G$ or, in a fancier language, a function annhilated
by the $n$th power of the augmentation ideal $I=\text{Ker}(\Z[\G]\to\Z)$. We make a few general remarks about these
higher order modular forms.  (For further properties and examples we refer the reader to
the recent preprint \cite{CDO}.)  First, if we denote by $M_k^{(n)}(\G)$ the vector space of $n$th order modular forms 
on $\G$, then $M_k^{(n)}(\G)$ is always finite-dimensional and in fact of dimension at most $(1+r+\cdots+r^{n-1})D$,
where $D=\dim M_k(\G)$ is the dimension of the space of ordinary modular forms of weight $k$ on $\G$ and $r$ is the
cardinality of a set of generators of $\G$.  Indeed, 
if $\a_1,\ldots,\a_r\in\G$ are generators, then the map sending $f$ to the $r$-tuple $(f|_k(\a_1-1),\ldots,f|_k(\a_r-1))$
maps $M_k^{(n)}(\G)$ to $(M_k^{(n-1)}(\G))^r$ and has kernel $M_k(\G)$, so the result follows by induction.  This bound,
however, is not sharp, since for instance for $\G=\G_1$ (the full modular group, generated by the elements $S$ and $U=ST$
of order 2 and 3), the space $M_k^{(n)}(\G)$ reduces to simply $M_k(\G)$ for all $n$, the reason being that 
$(1-S)^n=2^{n-1}(1-S)$ and $(1-U)^n(2+U)^{n-1}=3^{n-1}(1-U)$ in the group ring $\Z[\G_1]$, so that the equations $f|(1-S)^n=0$ 
and $f|(1-U)^n=0$ already imply $f|(1-S)=0$ and $f|(1-U)=0$.  The same argument applies to any group generated by elements 
of finite order. (In the situation studied here, the group involved {\it is} generated by elements of finite order, but 
the above argument can no longer be applied because of the presence of a character, i.e., because the diagonal terms in 
the two matrices in (\hoff) are roots of unity but are not all equal to~1.)  In general, however, the spaces $M_k^{(n)}(\G)$
are larger than $M_k(\G)$.  In particular, one can construct non-trivial modular forms of (say) second order and weight~2
simply by multiplying a modular form of weight~2 by the Eichler integral of another modular form of weight~2, as was done
for the function $f_2$.  Note, however, that $f_2$ is an atypical second order modular form since for a general such 
function $f$ we would only require that $f|(1-\g)$ is {\it some} modular form of weight~2 for each $\g\in\G$, while in 
the case of $f_2$ each of the functions $f_2|(\g-1)$ is a multiple of the {\it same} modular form $f_1$ (i.e., the
functions $f_1$ and $f_2$ together are the components of a vector-valued but first order modular form $F(\tau)$, as we 
saw above).

In summary, the study of a problem coming from statistical mechanics has led to the consideration of a new and
interesting type of modular object.

\subheading{8. Discussion}

\medskip  
{\bf a.} In this work, we examine the modular properties of crossing probabilities and their generalizations, 
considered as functions of $\tau = ir$, with $r$ complex. Consider, for definiteness, the horizontal crossing 
probability $\Pi_h(r)$. The original problem is defined on a rectangle, which corresponds to a fundamental domain 
of the modular lattice $\Z\tau+\Z$. However, the function defined by analytic continuation of  $\Pi_h(r)$ is {\it not} 
the correct crossing probability for percolation on the parallelogram generated by 1 and $\tau=ir$ when $r$ is 
not real. This holds for all the crossing probabilities and their derivatives, and is a basic manifestation 
of the problem of understanding the connection between the conformal and modular properties at play here (unlike the 
situation for, e.g., the partition function defined on a torus, where it is quite natural to expect modular invariance).  The physical quantity defined by Smirnov \cite{Sm} is in fact given by (the real part of) $\Pi_h(r)$ with $r$ complex, but it also does not appear to have any natural modular properties.  

Another way to describe this issue is via the ``mysterious" behavior of the crossing probabilities under the operation $T$, mentioned in the introduction.  In our treatment this behavior is introduced by the assumption of one conformal block in Theorems 1 - 3. In CFT and SLE, on the other hand, it arises from the particular differential equations that the crossing probabilities satisfy.
 
\medskip
{\bf b.} 
Note that $\Pi_h$ for percolation (for which $c = 0$) satisfies the condition that the boundary conditions are the same on both 
horizontal sides (see \cite{Ca2}) while $\Pi_{hv}$ does not, since the boundary conditions in that case are different
on all four sides of the rectangle \cite{W}.  This explains why the former is an even conformal block (see \cite{Ca4}), 
while the latter includes a function that is not.  (In fact, for $\Pi_h$, two conformal blocks appear in the CFT
calculation, but one of them is a constant, i.e. of dimension zero and with all coefficients vanishing except $a_0$.)  
On the other hand, the block for $\Pi_h(r;\a)$ for the SLE processes is not even except
for $\a = 0$, when it is a constant, or for $\a = 1/3$, i.e.~percolation, as a consequence of Theorem 1.  It follows from 
the arguments in \cite{Ca4} that this generalized crossing  probability cannot be expressed as a difference of partition 
functions with boundary conditions the same on both horizontal sides unless the dimensions of the conformal blocks that appear differ by 
half-odd integers, since each partition function would necessarily be expressible as a sum of even conformal blocks.

\medskip
{\bf c.} In some recent work, the plus-spin horizontal crossing function in the critical Ising model has been investigated 
numerically \cite{LLS}. This quantity satisfies (\dual) and the simulations indicate that it is conformally invariant and vanishes
as  $r\to\infty$ as $e^{-\a r}$ with $\a \approx 1/6$.  Although its vanishing is consistent with the asymptotic 
behavior of $\Pi_h(r;1/6)$, further work \cite{LSA} indicates that the agreement of $\Pi_h(r;1/6)$ with the numerical 
results is significantly worse than that of a particular Ising model CFT solution obtained by the authors.  This solution is 
not given by a single conformal block, which is consistent with Theorem 2 (if polynomial growth is assumed).

\subheading{Acknowledgements}

We acknowledge useful conversations and correspondence with M.~Aizenman, J.~L.~Cardy, R.~Kenyon, I.~Peschel, O.~Schramm,
W.~Werner, K.~Yasuda, and R.~M.~Ziff. One of us (PK) is grateful for the hospitality of the Max-Planck-Institut f\"ur 
Mathematik, Bonn, where part of this research was performed.  His research is based in part on work supported by the National Science Foundation under Grant No. DMR-0203589.

\end